\documentclass{appolb}
\usepackage{epsfig}

\begin{document}
\title{SCHWINGER TUNNELING AND THERMAL CHARACTER OF HADRON SPECTRA
%\thanks{Research supported in part by the Polish State Committee for
%        Scientific Research, grant 2 P03B 09419}%
}
\author{Wojciech Florkowski
\address{The H. Niewodnicza\'nski Institute of Nuclear Physics, \\
         Polish Academy of Sciences, \\
         31-342 Krak\'ow, Poland \\
        and \\
        Institute of Physics, \'Swi\c{e}tokrzyska Academy, \\
         25-406 Kielce, Poland\\}
}
\maketitle
\begin{abstract}
It is shown that an oscillatory character of the solutions of the
collisionless kinetic equations describing production of the
quark-gluon plasma in strong color fields leads to the exponential
(thermal-like) transverse-momentum spectra of partons produced in the
soft region ($100 \hbox{\,MeV} < p_\perp < 1 \hbox{\,GeV}$). In
addition, the production of partons in the very soft region ($p_\perp
< 100 \hbox{\,MeV}$) is clearly enhanced above the thermal-like
background.
\end{abstract}

\section{Introduction}

The transverse-momentum spectra of hadrons measured at RHIC are very
well reproduced by the thermal model \cite{wbwf}. Since the
thermal-like spectra appear also in the collisions of more elementary
systems, the questions arises if the thermal behavior observed at RHIC
may be truly attributed to the rescattering processes or is it of a
completely different origin connected, e.g., with a trivial
phase-space dominance effect (for a recent discussion of this and
similar issues see Refs. \cite{rischke,krzywicki,qmst,bim}).

In this paper we follow the idea formulated by Bialas and argue that
the thermal shape of the transverse-momentum spectra of hadrons may
have its origin in the fluctuations of the string tension.  In
Ref. \cite{AB} Bialas showed that the thermal character of the
measured transverse-momentum spectra,
\begin{equation}
{dN_{\rm exp} \over d^2p_\perp} \sim \exp\left(-m_\perp/T\right),
\quad m_\perp = \sqrt{m^2+p^2_\perp}\,,
\label{thermal}
\end{equation}
may be understood as an effect of the fluctuations of the string
tension $\kappa^2$ which appears in the Schwinger formula
\cite{schwinger,cnn,gm,gi},
\begin{equation}
{dN_{\rm Schwinger} \over d^2p_\perp} 
\sim \exp\left(-\pi m^2_\perp/\kappa^2\right).
\label{schwinger}
\end{equation}
Although the $m_\perp$-dependence in Eqs. (\ref{thermal}) and
(\ref{schwinger}) is different, the appropriate averaging of formula
(\ref{schwinger}) over $\kappa$ may produce indeed an exponential
function,
\begin{equation}
\int d\kappa \, P(\kappa) \, \exp\left(-\pi m^2_\perp/\kappa^2\right)
\sim \exp\left(-m_\perp/T\right).
\label{fluct}
\end{equation}
The explicit (gaussian) form of the distribution $P(\kappa)$ as well
as a relation connecting $T$ with the average value of $\kappa^2$,
\begin{equation}
T=\sqrt{\langle \kappa^2 \rangle \over 2 \pi},
\label{Tk}
\end{equation}
was given and discussed in Ref. \cite{AB}.

In this paper we show that the situation described above appears
naturally in the kinetic equations describing production of the
quark-gluon plasma in strong color fields. In this case, due to the
screening effects, the color fields change in time and may even
oscillate \cite{BCDFosc}. As a consequence, the transverse-momentum
spectra acquire a form very similar to Eq. (\ref{fluct}).  The only
difference is that $\kappa^2$ should be treated now as a function of
time
\begin{equation}
\int dt \, P^{\,\prime}(t) \, \exp\left(-\pi m^2_\perp/\kappa^2(t)\right)
\sim \exp\left(-m_\perp/T\right).
\label{tfluct}
\end{equation}
The form of the distribution $P^{\,\prime}(t)$ is uniquely determined
by the kinetic equations and, as we shall see, formula (\ref{tfluct})
yields effectively the exponential spectra in the soft region, $100
\hbox{\, MeV}< p_\perp < 1 \hbox{\, GeV}$.  For larger values of
$p_\perp$ the model based on the Schwinger formula gives the spectrum
which decays faster than the exponential function. However, in this region
the production of particles becomes a hard process and the
use of the Schwinger formula is inadequate.  On the other hand, for
very small values of $p_\perp$ we find an enhancement above the
exponential background, which is a desirable effect in view of the
experimental measurements of the pion spectra which consistently show
such an increase.

Although there is a formal similarity between our formulas and those
used by Bialas, there is also an important physical difference between
the two approaches.  In our calculations we consider the values of the
string tension which are larger than the elementary string
tension. This may be a realistic situation in heavy-ion collisions
\cite{bironk}. On the other hand, Bialas considers possible
fluctuations of the elementary string tension, which may appear due to
stochastic nature of the QCD vacuum \cite{heid}. Thus, our approach
may explain the origin of the thermal spectra observed in heavy-ions
but it is not capable of describing the thermal features observed in,
e.g., electron-positron annihilations. It is conceivable, however,
that the effect of the stochastic vacuum plays an additional role in
the heavy-ion collisions leading to even more pronounced
thermalization effects.

\section{Tunneling of partons in oscillatory chromoelectric fields}

In our approach we use the semi-classical kinetic equations for the
quark-gluon plasma written in the abelian dominance approximation
\cite{BCDFosc,heinz,egv,BCapp,DFhq}
\begin{equation}
\left( p^{\mu }\partial _{\mu } \pm g{\mbox{\boldmath $\epsilon$}}_{i}\cdot 
{\bf F}^{\mu \nu }p_{\nu }\partial _{\mu }^{p}\right) 
G^\pm_{i}(x,p)=\frac{dN^\pm_{i}}{d\Gamma },  \label{kineq}
\end{equation}
\begin{equation}
\left( p^{\mu }\partial _{\mu }+g{\mbox{\boldmath $\eta$}}_{ij}\cdot 
{\bf F}^{\mu \nu }p_{\nu }\partial _{\mu }^{p}\right) 
\tilde{{G}}_{ij}(x,p)=
\frac{d{\tilde{N}}_{ij}}{d\Gamma },  \label{kineg}
\end{equation}
where $G^+_{i}(x,p)\ $, $G^-_{i}(x,p)$ and $\tilde{{G}}_{ij}(x,p)$ are
the phase-space densities of quarks, antiquarks and gluons,
respectively.  Here $g$ is the strong coupling constant and
$i,j=(1,2,3)$ are color indices. The terms on the left-hand-side
describe the free motion of the particles as well as their interaction
with the mean color field $\mathbf{F}_{\mu \nu }$. The terms on the
right-hand-side describe production of quarks and gluons due to the
decay of the field.  In our present calculations we include only the
two lightest flavors and neglect the quark masses ($m_\perp=p_\perp$).

We note that Eqs. (\ref{kineq}) and (\ref{kineg}) do not include any
thermalization effects. The latter can be taken into account if the
collision integrals are incorporated on the right-hand-side of
Eqs. (\ref{kineq}) and (\ref{kineg}). So far, most of the approaches
have included the collision integrals in the relaxation-time
approximation \cite{bbr,bn,brs}. A more recent and elaborated
treatment of the collision integrals may be found in Ref. \cite{mrow}.
We note also that the semi-classical kinetic equations may be derived
within a field-theoretic approach if a separation of different time
scales can be achieved: the time scales associated with quantum phase
oscillations and amplitudes of pair creation should be much smaller
than the time scales associated with the oscillations of the fields
\cite{cm,kescm,cekms,kme}.

In the next Sections we shall consider a one-dimensional (i.e.,
uniform in the transverse direction) boost-invariant system. In this
case it is convenient to use the boost-invariant variables introduced
in Refs. \cite{BCprd}
\begin{equation}
u=\tau ^{2}=t^{2}-z^{2},\quad w=tp_{\Vert }-zE,\quad \mathbf{p}_{\bot },
\label{binvv1}
\end{equation}
and also 
\begin{equation}
v=Et-p_{\Vert }\ z=\sqrt{w^{2}+m_{\perp }^{2}u}.  \label{binvv2}
\end{equation}
From these two equations one can easily find the energy and the longitudinal
momentum of a particle 
\begin{equation}
E=p^{0}=\frac{vt+wz}{u}=p_\perp \cosh y,\quad p_{\Vert
}=\frac{wt+vz}{u}
=p_\perp \sinh y.  \label{binvv3}
\end{equation}
Besides the rapidity $y$, we also introduce the quasirapidity variable
$\eta$ which is related to the space-time coordinates $t$ and $z$ by
equations
\begin{equation}
t= \tau \cosh\eta, \,\,\,\, z=\tau \sinh\eta.
\label{eta}
\end{equation}
The invariant measure in the momentum space is
\begin{equation}
dP = d^2p_\perp {dp_\parallel \over p^0} = d^2p_\perp {dw \over v},
\label{dP}
\end{equation}
whereas in the Minkowski space-time the appropriate measure has the
form
\begin{equation}
d^4x = \tau \sinh\eta\,d\tau\,d\eta\,dx\,dy.
\label{d4x}
\end{equation}
The invariant measure in the phase-space is $d\Gamma = d^4x d^3p/p^0$.
In the considered situation, the only non-zero components of the
tensor ${\bf F}_{\mu \nu }=(F_{\mu \nu }^{3},F_{\mu \nu }^{8})$ are
those corresponding to the color field ${\mbox{\boldmath $\cal
E$}}={\bf F}^{30}$.  The quarks and gluons couple to the field
${\mbox{\boldmath $\cal E$}}$ through the charges ${\mbox{\boldmath
$\epsilon$}}_{i}$ and ${\mbox{\boldmath $\eta$}}_{ij}$ defined in
\cite{BCDFosc,huang}.

\section{Transverse-momentum spectra}

For one-dimensional boost-invariant systems the production rates
appearing on the right-hand-side of Eqs. (\ref{kineq}) and
(\ref{kineg}) have a general form \cite{bajan}\footnote{ We neglect
here the finite-size effects in the pseudorapidity space taken into
account in Ref. \cite{bajan}, since they have
a negligible effect on the time evolution of the system.}
\begin{equation}
\frac{dN}{d\Gamma }
= p^{0}\frac{dN}{d^{4}x\ d^{3}p}=\frac{F}{4\pi ^{3}}\left|
\ln \left( 1\mp \exp \left( -\frac{\pi p_{\perp}^{2}}{F}\right) \right)
\right| \delta \left( w-w_{0}\right) v,  \label{rate1}
\end{equation}
where $F$ is the force acting on a parton (for the boost-invariant
systems $F$ depends only on $\tau$ and the color charge of a quark or
a gluon), $w_{0}$ is the longitudinal momentum gained by a parton
during the tunneling process \cite{bajan,BCDFtun},
\begin{equation}
w_{0} =-\frac{ p_{\perp }^{2}}
{2F},
\label{wf}
\end{equation}
and the plus/minus sign is connected with the statistics of the
tunneling particles (plus for bosons and minus for
fermions). Introducing the notation
\begin{equation}
\frac{dN}{d\Gamma } =  {\cal R}(\tau,p_\perp)
\delta \left( w \mp w_{0}\right) v,
\label{qrate}
\end{equation}
we find that the transverse-momentum spectra of partons are
given by the formula
\begin{eqnarray}
\frac{dN}{dy\, d^2p_\perp } 
&=& \int d^4x \frac{dN}{d\Gamma } = \pi R^2 \int\limits_0^\infty
d\tau^\prime \, \tau^\prime \int\limits_{-\infty}^{+\infty} d\eta \,
{\cal R}(\tau^\prime,p_\perp) \delta \left( w \mp w_{0}\right) v 
\nonumber \\
&=&  \pi R^2 \int\limits_0^\infty
d\tau^\prime \, \tau^\prime \,{\cal R}(\tau^\prime,p_\perp), 
\end{eqnarray}
or more explicitly
\begin{equation}
\frac{dN}{dy\, d^2p_\perp } =
{ R^2 \over 4 \pi^2} \, \sum_{\rm all\,\, partons} \, 
\int\limits_0^\infty 
d\tau^\prime \, \tau^\prime  \, F(\tau^\prime)
\left|
\ln \left( 1\mp \exp 
\left( -\frac{\pi p_{\perp }^{2}}{F(\tau^\prime)}\right) \right)
\right|.
\label{main}
\end{equation}
Here we have introduced the sum over all tunneling partons, i.e.,
quarks and gluons, and $R$ is the transverse radius of our system.
However, for simplicity of notation we skip the indices denoting
different quantum numbers of partons. The explicit expressions for $F$
and all other details are given in Ref. \cite{bajan}. In the numerical
calculations we use the value $\pi R^2 = 1 \, \hbox{fm}^2$, hence our
results describe the production of partons per unit transverse
area. Eq. (\ref{main}) is the counterpart of Eq.  (\ref{fluct})
studied by Bialas.

We note that formula (\ref{main}) may be alternatively obtained from
the Cooper-Frye formalism outlined shortly in the Appendix. We also
note that the transverse-momentum spectra have not been calculated so
far in the formalism outlined in Sect. 2, only the mean $p_\perp$ was 
studied in Ref. \cite{BCapp}.

\section{Results}

The starting point of our calculation is Eq. (\ref{main}).  The time
dependence of the forces $F$ is known (in the numerical form) from the
studies performed in Ref. \cite{bajan}. In practice, the integration
range over $\tau^\prime$ is always finite; the forces $F(\tau^\prime)$
are different from zero only at the initial stage of the evolution of
the system ($\tau^\prime < 1.5$ fm).  The initial condition for the
color field is obtained from the Gauss law

\begin{equation}
{\mbox{\boldmath $\cal E$}}_{0}{\,=}
\sqrt{\frac{2\sigma _{g}}{\pi R^{2}}}\,k\,
{\mbox{\boldmath $\eta$}}_{12}.  
\label{initcon}
\end{equation}
Here the string tension $\sigma_g= 3 \sigma_q =$ 3 GeV/fm, and the
number of color charges which span the initial field is denoted by $k$
(note that ${\mbox{\boldmath $\eta$}}_{12}=(1,0)$).

In Figs. 1 and 2 we show our results obtained for $k=2$ and $k=3$.
The transverse-momentum spectra are represented by the solid lines.
In both cases one can observe that the spectra may be well
approximated by the exponential function, especially in the soft
region $100 \hbox{\,MeV} < p_\perp < 1 \hbox{\,GeV}$. Hence, the
Schwinger tunneling mechanism in time-dependent fields indeed leads to
the thermal-like spectra, although no rescattering processes are taken
into account in this picture.  

The dashed lines in Figs. 1 and 2 represent the exponential functions
with the inverse-slope parameter $T$ and the normalization fitted at
$p_\perp$ = 350 MeV.  The inverse-slope parameters are $T=220$ MeV and
$T=270$ for $k=2$ and $k=3$, respectively. We thus see that larger
initial fields lead to higher effective temperatures. This is already
expected from Eq. (\ref{Tk}), since larger initial fields lead to
larger fluctuations (changes) of the field in time and, finally, to
larger values of the parameter $T$. We may even try to apply
Eq. (\ref{Tk}) in our case, replacing the average value of $\kappa^2$
by the time average of $F$. In this way we find $T=258$ MeV for $k=2$
and $T=307$ MeV for $k=3$. As we can see, the rough estimate based on
Eq. (\ref{Tk}) gives the correct magnitude of $T$.  In the case $k=2$
we find the mean transverse momentum $\langle p_\perp \rangle$ = 376
MeV and the rapidity density (per unit transverse area) $dN/dy$ =
1.2. In the case $k=3$ we find $\langle p_\perp \rangle$ = 426 MeV and
$dN/dy$ = 1.9. These results are consistent with the earlier reported
values \cite{DFhq}

\begin{figure}[h]
\epsfysize=6cm
\par
\begin{center}
\mbox{\hspace{-1cm} \epsfbox{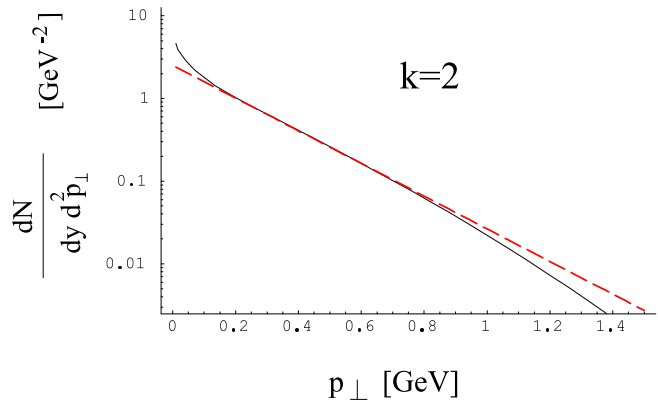}}
\end{center}
\caption{The transverse-momentum spectrum of quarks and gluons
obtained in the case $k=2$ (solid line) and the exponential function
const$\times\exp[-p_\perp/(220 \hbox{MeV})]$ \, (dashed line).  The
inverse slope parameter $T$ was fitted at $p_\perp$ = 350 MeV. }
\label{k2}
\end{figure}

\begin{figure}[h]
\epsfysize=6cm
\par
\begin{center}
\mbox{\hspace{-1cm} \epsfbox{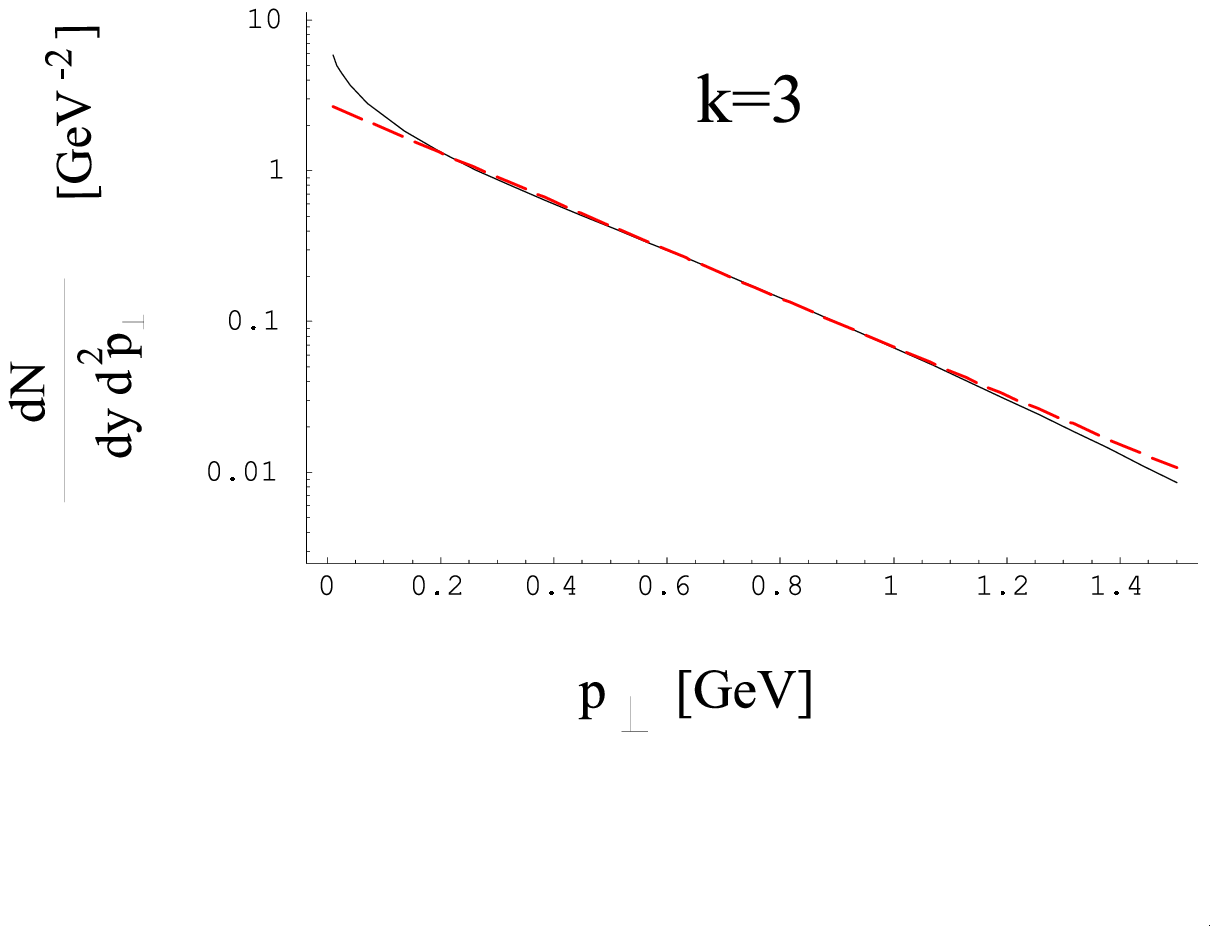}}
\end{center}
\caption{The spectrum obtained in the case $k=3$ (solid line) and the
exponential function const$\times\exp[-p_\perp/(270 \hbox{MeV})]$ \,
(dashed line).  The inverse slope parameter $T$ was fitted also at
$p_\perp$ = 350 MeV.}
\label{k3}
\end{figure}

Another interesting feature of the spectra shown in Figs. 1 and 2 is
the enhancement of the particle production above the thermal-like
background in the very soft region $p_\perp <$ 100 MeV. This type of
the behavior is observed in the pion spectra measured by various
experiments at CERN and RHIC, and is usually explained as the effect
of the resonance decays which give contributions mainly in the
low-$p_\perp$ region.  The production of such very soft partons
occurs in our model at later times when the forces $F$ are small and
the production of the particles with large $p_\perp$ is strongly
suppressed. In other words, this may be treated as a phase-space
effect -- when the initial string breaks into many small pieces, only
particles with small $p_\perp$ can tunnel and they contribute mainly
to the low-$p_\perp$ peak. Nota bene, such type of the behavior was
also found in the simulations of the sequential decays of the
color-flux tubes \cite{DFsim}.

We conclude that the Schwinger tunneling mechanism in strong varying
fields offers an appealing explanation of a very fast formation of the
thermal-like system in heavy-ion collisions.

\section{Appendix: Cooper-Frye formula}

The transverse-momentum spectra may be calculated from the Cooper-Frye
formula \cite{CF}
\begin{equation}
\frac{dN}{dy\, d^2p_\perp } = \int d\Sigma_\mu(x) p^\mu \,
f(x,p).
\label{CF}
\end{equation}
In Eq. (\ref{CF}) the quantity $d\Sigma_\mu(x)$ is the element of the
freeze-out hypersurface and $f(x,p)$ denotes the phase-space
distribution function. Assuming that the system is boost invariant in
the longitudinal $(z)$ direction and uniform in the transverse $(x,y)$
directions, we may rewrite Eq. (\ref{CF}) in the form
\begin{equation}
\frac{dN}{dy\, d^2p_\perp } = \int dx \,dy \,d\eta \,v 
\,f(\tau,w,p_\perp).
\label{CFbi}
\end{equation}
Here we used the property $d\Sigma_\mu=u_\mu \tau \,d\eta \,dx \,dy$,
which follows from the condition that freeze-out occurs at a constant
value of the invariant time $\tau$. We also used the explicit form of
the boost-invariant four-velocity, $u^\mu=(t,0,0,z)/\tau$, which gives
$p^\mu u_\mu= v/\tau.$ Now using the explicit form for the solutions
of the kinetic equations (\ref{kineq}) and (\ref{kineg}) obtained in
Ref. \cite{bajan}
\begin{equation}
f(\tau,w,p_\perp)=\int\limits_0^\tau \,d\tau^\prime \,\tau^\prime
{\cal R}(\tau^\prime,p_\perp) \delta\left( \Delta h(\tau,\tau^\prime)
\pm w - w_0(\tau^\prime,p_\perp) \right)
\label{solu}
\end{equation}
we arrive at formula (\ref{main}).

\end{document}